# Superlinear density dependence of singlet fission rate in tetracene films

Bo Zhang, [1] Chunfeng Zhang, [1,*] Rui Wang,[1] Zhanao Tan,[2] Yunlong Liu, [1] Wei Guo,[1] Xiaoling Zhai,[1] Xiaoyong Wang, [1] and Min Xiao[1,3,†]

[1] National Laboratory of Solid State Microstructures, Department of Physics, Nanjing University, Nanjing 210093, China

[2] New and Renewable Energy of Beijing Key Laboratory, School of Renewable Energy, North China Electric Power University, Beijing 102206, China

[3] Department of Physics, University of Arkansas, Fayetteville, Arkansas 72701, USA

**Abstract:**

We experimentally show that the rate of singlet fission in tetracene films has a superlinear dependence on the density of photo-excited singlet excitons with ultrafast transient absorption spectroscopy. The spectrotemporal features of singlet and triplet dynamics can be disentangled from experimental data with the algorithm of singular value decomposition. The correlation between their temporal dynamics indicates a nonlinear density dependence of fission rate, which leads to a conjecture of coherent singlet fission process arising from superradiant excitons in crystalline tetracene. This hypothesis might be able to resolve some long-standing controversies.

Singlet fission in organic semiconductors is a spin-allowed process that creates two triplet excitons ($2T_1$) from one photo-excited singlet exciton ($S_1$) [1-7]. This process has been proposed to beat the Shockley-Queisser theoretical limit [8] in single-junction solar cells [9, 10]. Very recently, its promising potential has been evidenced by the rapid development of singlet-fission-sensitized solar cells and photodetectors [11-15]. However, the exact physical mechanism for singlet fission, particularly in a prototypical material of tetracene, is still under intensive debate [3, 4, 7, 16-19]. Some new perspectives inspired by recent experiments have put forward interesting concepts such as quantum coherence [20, 21], electron correlation [22], and intermolecular interaction [17, 23], in addition to the conventional charge-transfer mechanism [5].

The difficulty in elucidating singlet fission process in tetracene is partially caused by the endothermic nature ($E_{S_1} - 2E_{T_1} < 0$) [1, 3, 16, 18, 19, 24-28]. Thermal activation was originally proposed for compensating the energy uphill [1, 3, 26-28]. This argument was supported by an ultrafast transient absorption (TA) study, in which Thorsmølle *et al.* observed a significant temperature dependence of the triplet signal in tetracene single crystals [3]. However, opposite conclusions have been drawn in recent studies on tetracene films [18, 19]. In particular, with a similar TA experiment, Wilson *et al.* reported that the singlet fission process is actually independent of temperature [19]. In addition, the measured singlet-fission rate in crystalline tetracene has also been an unresolved issue in literature. The singlet fission process has usually been phenomenologically described by using a rate equation as $dN_T(t)/dt = 2k_{SF}N_S(t)$ [16, 24, 29], where $k_{SF}$ is a rate constant that connects the dynamics of singlet ($N_S$) and triplet ($N_T$) populations. However, the deduced values of $k_{SF}$ in early experiments varied broadly in a range of $10^8$-$10^{12}$ s$^{-1}$ [5]. The ultrafast time-resolved (TR) spectroscopy should have provided much converging results for the singlet-fission rate. Nevertheless, there is still over one-order divergence in its value among the latest reports ($10^{10}$ s$^{-1}$-$10^{11}$ s$^{-1}$) [18, 19, 24].

To address these outstanding controversies, we have systematically examined the density dependence of singlet fission process in tetracene films with broadband ultrafast TA spectroscopy. We extract the spectrotemporal features of singlet and triplet excitons from the TA data using the algorithm of singular value decomposition (SVD). The constructed correlations between their dynamics show compelling evidences that instead of a constant value, $k_{SF}$ is strongly dependent on the singlet exciton density, namely, the rate of singlet fission has a superlinear dependence on the density of photo-excited singlet excitons. In analogue to the process of superradiation, we conjecture that photo-generated singlet excitons in tetracene films may undergo a coherent fission process (the nonlinear part) in addition to the conventional spontaneous fission process (the linear part). The coherent singlet fission may arise from the superradiant excitons that interact with light in a collective and coherent fashion. This hypothesis can well explain the debated issues mentioned above in the studies of singlet fission in tetracene.

We prepared the tetracene films with thickness of ~ 100 nm by evaporating tetracene powders thermally on pre-cleaned glass substrates. The lowest-lying exciton states ($S_1$) in the samples are characterized with two peaks at 2.34 eV and 2.46 eV in the absorption spectrum due to the Davydov splitting [30]. The presence of singlet fission has been confirmed by performing TR-fluorescence (TRFL) spectroscopy [31] [Supplementary Information]. We utilized a commercial Ti:Sapphire regenerative amplifier (Libra, Coherent) at 800 nm with a repetition rate of 1 kHz and pulse duration of ~ 90 fs to carry out the TA experiments. For broadband measurements, a second-harmonic light source at 400 nm was used as the pump beam and an optical parametric amplifier (OperA solo, Coherent) was used to provide a probe beam with tunable wavelength. The relative polarizations of the pump and probe beams were set to be at the magic angle (54.7°). We used balance detection scheme together with a lock-in amplifier to measure the fractional change in transmission (ΔT/T) with a sensitivity better than $10^{-5}$. The experiments were performed with samples in vacuum provided by a cryostat (MicroCryostatHe, Oxford).

The measured ΔT/T in the 1.3-2.6 eV spectral region within a temporal window of 3 ns monitors the dynamics of singlet-exciton fission process [Fig. 1(a)]. The excitation flux was kept at a low level of ~ 5 μJ/cm$^2$ to minimize the effect of bimolecular recombination. The transient optical response consists of multiple entangled components due to superposition of stimulated emission, ground-state bleaching, excited-state absorption and so on. In tetracene, the transitions of $S_0 \rightarrow S_1$, $S_1 \rightarrow S_n$ and $T_1 \rightarrow T_n$ are probably responsible for the measured TA signal [3, 16, 19]. However, it is challenging to exactly assign the transient features at single wavelengths to specific transitions since the complicated structures of $S_n$ and $T_n$ are poorly understood. In the current study, the major bands of photo-induced bleaching and absorption locate at ~ 2.3 eV and ~ 1.8 eV [Fig. 1(a)], respectively. Their instant build-up processes as well as the fast decays at early stage [Figs. 1(b) & 1(c)] are the characteristics of singlet excitons [3, 19]. Nevertheless, the long-lived tails, with the amplitude ratios much larger than that exhibited in the TRFL curves, imply that the triplet excitons may also make some contributions.

The spectrotemporal features for the triplet excitons are more complicated. In general, triplet excitons are characterized by long-lived signals of photo-induced absorption. Such features have been reported with probe energies at 1.67 eV [3, 19], 2.53 eV [4] and 2.67 eV [16] in different studies, respectively. Here, a similar long-lived band of photo-induced absorption appears at ~ 1.60 eV. This component becomes prominent after ~ 100 ps and remains to be largely static in the time scale between 0.1-3 ns. The build-up process consists of one sub-picosecond component and one delayed-rising component [Fig. 1(d)]. Thorsmølle *et al.* assigned the two components to exciton fission from higher ($S_n \rightarrow 2T_1$) and lowest singlet states ($S_1 \rightarrow 2T_1$), respectively [3]. Wilson *et al.* proposed an alternative explanation that the sub-picosecond rise is also contributed by singlet excitons [19]. In this scenario, the signal at this band can be disentangled into two parts as

$$\Delta T(\text{t}) / T \propto A_S N_S(\text{t}) + A_T N_T(\text{t}), \quad (1)$$

where $A_S$ and $A_T$ are the constants in proportion to the oscillation strengths of singlet and triplet excitons, respectively; $N_S$ includes the densities of free and trapped singlet excitons; and $N_T$ is the triplet density including the correlated triplet pairs and separate triplet excitons.

We shifted the photon energy of pump beam to 2.3 eV for near-resonant excitation of $S_0 \rightarrow S_1$ to minimize the possibility of $S_n \rightarrow 2T_1$. In comparison to the excitation at 3.1 eV (400 nm), no significant change in the amplitude ratio of the sub-picosecond rise component has been observed [Figs. 2a & 2b], which indicates that the contribution made by $S_n \rightarrow 2T_1$ is unimportant, and thus it is reliable to ascribe the fast growth part to singlet excitons [19, 20]. This is reasonable since the relaxation of $S_n \rightarrow S_1$ is very fast considering the strong electron-phonon coupling in organic semiconductor crystals [32-34].

The assignment of single wavelength data to a specific transition is not accurate due to the entanglement of multiple transitions. To address this issue, we have employed the established algorithm of SVD [35, 36] to disentangle the spectrotemporal features of singlet and triplet excitons. The detailed description of SVD is available in Supplementary Information. Basically, one can decompose the two-dimensional matrix of the signal as $\Delta T / T(E, t)$ by calculating a corresponding matrix, $\Sigma_{k=1}^{l} \phi_k(E) \varphi_k(t)$, that best reproduces the experimental data with $\phi_k(E)$ and $\varphi_k(t)$ as the spectral and temporal eigenfunctions, respectively. The spectrotemporal features for the singlet and triplet excitons are reconstructed as a linear combination of these eigenfunctions. This method has been successfully used in extracting both energy- and time-domain information from a set of data in several research areas [35-38]. Here, we approximately adopt the signals at $t \sim 0.5 \, ps$ and $t \sim 3 \, ns$ as the spectral features to extract the temporal dynamics of the singlet and triplet excitons, respectively [39]. Such SVD approach has well reproduced the experimental data as compared in Figs. 3(a) & 3(b).

The constructed temporal dynamics for singlet and triplet excitons are shown in Fig.3(c). We use the phenomenological stretched-exponential (SE) equations in the forms of $N_S(t) = N_{S0}e^{-(t/\tau_S)^{\beta_S}}$ and $N_T(t) = N_{T0}(1 - e^{-(t/\tau_T)^{\beta_T}})$ to fit the decay and growth dynamics of singlet and triplet excitons, respectively. The fitted parameters are $\tau_S \sim 27.2$ ps, $\beta_S \sim 0.70$, $\tau_T \sim 21.6$ ps, and $\beta_T \sim 0.81$, respectively. In the following, we exam the density dependence of singlet fission by analyzing the correlation between the dynamics of singlet and triplet excitons. If the spin-forbidden intersystem crossing ($S_1 \rightarrow T_1$) is neglected, the dynamics of triplet population can be expressed in the rate equation as [16, 19]

$$dN_T(t)/dt = 2k_{SF}N_S(t) - k_{SP}^T N_T(t) - k_{TT}N_T^2(t) \text{ , (2)}$$

where $k_{SF}$, $k_{SP}^T$, and $k_{TT}$ are the rate constants for the processes of singlet fission, spontaneous recombination, and triplet-triplet annihilation, respectively. Since the latter two processes are very slow, only the process of singlet fission is responsible for the early-stage dynamics ($t < 100\,ps$), i.e. $dN_T(t)/dt \approx 2k_{SF}N_S(t)$. Let's employ a mathematical approximation to compare the values of $dN_T(t)/dt$ with $N_S(t)$ because the direct differentiation of $dN_T(t)/dt$ from the experimental data is quite noisy. For a SE growth dynamics, the differential value $dN_T(t)/dt = \dfrac{N_{T0}\beta_T t^{\beta_T-1}}{\tau_T^{\beta_T}}e^{-(t/\tau_T)^{\beta_T}}$ can be approximated as

$$dN_T(t)/dt \approx \frac{N_{T0}}{\tau_T}e^{-(t/\tau_T)^{\beta_T}} \propto N_{T0} - N_T(t) \text{ , (3)}$$

when $\beta_T \to 1$. It is a good approximation for $\beta_T \to 0.81$ [inset, Fig. 4(a)]. The curve of $N_{T0} - N_T(t)$ is plotted in a normalized scale together with $N_S(t)$ and $N_S^2(t)$ in Fig. 4(a). Surprisingly, $N_{T0} - N_T(t)$ curve shows a significant disparity from the linear dependence ($N_S(t)$), but close to the quadratic dependence ($N_S^2(t)$), when $t < 10\,ps$. This result strongly suggests that, rather than being a constant, $k_{SF}$ is actually density dependent, i.e. $k_{SF} = k_{SF0}(1 + \alpha N_S)$. That is, the rate of singlet fission is superlinearly dependent on the density of singlet excitons. For an accurate analysis, we rewrite the triplet dynamics in an integration form, i.e.

$$N_T(t) = \int_0^t 2k_{SF}N_S(\tau)d\tau. \qquad (4)$$

The temporal evolution behavior of the triplet exciton population [Fig. 4(b)] can be well fitted by the superlinear dependence with the parameter of $\alpha N_{S0} \approx 0.7$. We have further checked this model and found that the superlinear density dependence (Eq. 4) can well reproduce the measured results (< 100 ps) with excitation power up to 200 μJ/cm$^2$.

Such deduced density-dependent fission rate can resolve some of the controversies currently outstanding in various previous studies. The very different $k_{SF}$ values obtained in those studies can now be easily ascribed as having different carrier densities. We also notice that the fission yields in those previous studies were claimed to be efficient in spite of having such diverse densities [5, 6], which cannot be explained if $k_{SF}$ were really density independent considering the competing channels of singlet-singlet annihilation and superradiation. However, this fact can be naturally explained if one assumes a superlinear density dependence for the fission rate. In general, the rate equation for the relaxation of singlet population can be expressed as [16, 19]

$$dN_S / dt = -k_{SF} N_S - k_{SP}^S N_S - k_{SR}^S N_S^2 - k_{SS} N_S^2 + c k_{TT} N_T^2, \qquad (5)$$

where $k_{SP}^S$, $k_{SR}^S$ and $k_{SS}$ are the rate constants for spontaneous recombination (including spontaneous emission and nonradiative recombination like defect trapping), superradiation, and singlet-singlet annihilation, respectively. The constant $c$ (in the range of 0~1) takes into account the generation of singlet excitons through triplet-triplet annihilation [40], which is much less efficient in comparison to other processes. The quantum yield for singlet fission can then be estimated as

$$\eta_{SF} = \frac{k_{SF0} + \alpha N_S}{k_{SF0} + k_{SP}^S + (\alpha + k_{SR}^S + k_{SS}) N_S}. \qquad (6)$$

Approximately, $\eta_{SF}$ approaches to $k_{SF0} / (k_{SF0} + k_{SP}^S)$ at the low density limit and $\alpha / (\alpha + k_{SR}^S + k_{SS})$ at the high density limit, respectively. From this one can conclude that efficient singlet fission can occur in crystalline tetracene even in the high density regime where bimolecular recombination becomes dominant.

The intrinsic physics underlying the superlinear density dependence is yet to be elucidated. Here, we present a qualitative explanation concerning the superradiant state of singlet excitons in tetracene. The dependence of fission rate on the exciton density is similar to that of certain coherent processes like stimulated emission/absorption/scattering and superradiation, which leads us to conjecture that coherence processes between singlet excitons may also be involved in singlet fission here. In crystal tetracene, only the coherent process of superradiation has been demonstrated [41-43]. In a TRFL study, it's been argued that the  excitons can delocalize in a scale of ~ 10 molecules [41]. These superradiant excitons, whose transition dipoles interact with light in a collective and coherent fashion [43], may also undergo exciton fission coherently, which can manifest itself as the nonlinear part of the fission rate, i.e. $\alpha N_S^2$ [44]. The energy uphill ($E_{S_1} - 2E_{T_1}$) in such system with multiple coherently-correlated excitons may not be so significant as encountered in the case of single molecules. The reduced endothermicity can be easily compensated by certain intrinsic effects like intermolecular interactions and/or electronic coupling as previously proposed in literatures [17, 18, 22]. The thermal activation is no longer a prerequisite for the process of coherent exciton fission. In other words, the photo-excited singlet excitons undergo two types of fission processes: (1) the coherent fission process arising from the superradiant excitons that interact with light coherently; and (2) the spontaneous fission process similar to that in exothermic systems such as pentacene and its derivatives [2, 3, 29, 45, 46], for which thermal activation is required in crystalline tetracene to overcome the energy uphill.

Next, let's understand the sharp divergence on the issue of thermal activation argued in the previous works. In the work of Wilson *et al*, the fission rate in tetracene films was reported to be nearly temperature independent after subtracting off the singlet contribution from the long-lived triplet signal [19]. In the work of Thorsmølle *et al*, the triplet signature (~ 1.67 eV) at ~ 200 ps drops to half when cooling to 150 K in single crystals [3]. The temperature dependence is still distinct even if the fast rise component has been subtracted according to the approach in Ref. [19] (Eq.1), because most of the singlet excitons recombine before 200 ps. Note also that the role of thermal activation has only been highlighted in single crystals [3, 25] but not in films [16, 18, 19, 24]. These results suggest that the crystalline quality may be important. In tetracene films, singlet excitons decay much faster while the magnitude ratio between triplet and singlet signals is much smaller [16, 24]. The trapping effect leads to a much faster spontaneous recombination ($k_{SP}^S$) in thin films where the densities of trapping centers (e.g. defects and grain boundaries) are high, leading to a very low efficiency of spontaneous exciton fission. In this case, only triplet population induced by the coherent fission process (i.e. the nonlinear part) can be detected, which is insensitive to the temperature. On the other hand, in addition to the coherent fission process, the spontaneous fission process in single crystals is probably also efficient since thermal activation at room temperature is sufficient for compensating the energy uphill. To check this, we have performed the experiments on tetracene single crystals and analyzed the room-temperature data with the SVD algorithm [Supplementary Information]. The value of $k_{SF}$ in single crystal is much closer to a linear density dependence compared to the data from tetracene films, which confirms the relatively efficient spontaneous fission process in tetracene single crystals. The part of triplet signal in this case can be suppressed by cooling the sample [3].

In summary, we have analyzed the ultrafast TA data with the SVD algorithm to elucidate the dynamics of singlet-exciton fission in tetracene films. The observed superlinear density-dependent fission rate leads to a hypothesis of the coherent fission process in analogue to the superradiance. Theoretically, the coherent fission process may be further studied by describing the superradiant excitons with the Dicke model in future [44]. The results in this work provide comprehensive explanations to the debated issues of divergence in measured fission rates and thermal activation mechanism in crystalline tetracene, which should have significant impacts for studying singlet fission in other systems. The indication of the key role played by certain coherent behavior between delocalized excitons for singlet fission can be used to guide the search for singlet-fission sensitizers to improve performance of solar cells and photodetectors.

This work is supported by the National Basic Research Program of China (2013CB932903 and 2012CB921801, MOST), the National Science Foundation of China (91233103, 61108001, 11227406 and 11321063), and the Priority Academic Program Development of Jiangsu Higher Education Institutions (PAPD). We acknowledge Dr Vitaly Podzorov, Dr Yuanzhen Chen, and Dr Jonathan Burdett for providing valuable information for crystal growth and sample characterization.

# References and notes

E-mail: *cfzhang@nju.edu.cn (C.Z.), †mxiao@uark.edu (M.X.)

[1]   Geacinto.N, M. Pope, and F. Vogel, Phys. Rev. Lett. 22, 593 (1969).

[2]   H. Marciniak, M. Fiebig, M. Huth, S. Schiefer, B. Nickel, F. Selmaier, and S. Lochbrunner, Phys. Rev. Lett. 99, 176402 (2007).

[3]   V. K. Thorsmolle *et al.*, Phys. Rev. Lett. 102, 017401 (2009).

[4]   E. M. Grumstrup, J. C. Johnson, and N. H. Damrauer, Phys. Rev. Lett. 105, 257403 (2010).

[5]   M. B. Smith and J. Michl, Chem. Rev. 110, 6891 (2010).

[6]   M. B. Smith and J. Michl, Annu. Rev. Phys. Chem. 64, 361 (2013).

[7]   D. Beljonne, H. Yamagata, J. L. Bredas, F. C. Spano, and Y. Olivier, Phys. Rev. Lett. 110, 226402 (2013).

[8]   W. Shockley and H. J. Queisser, J. Appl. Phys. 32, 510 (1961).

[9]   M. C. Hanna and A. J. Nozik, J. Appl. Phys. 100, 074510 (2006).

[10] I. Paci, J. C. Johnson, X. Chen, G. Rana, D. Popovic, D. E. David, A. J. Nozik, M. A. Ratner, and J. Michl, J. Am. Chem. Soc. 128, 16546 (2006).

[11] B. Ehrler, M. W. B. Wilson, A. Rao, R. H. Friend, and N. C. Greenham, Nano Lett. 12, 1053 (2012).

[12] P. J. Jadhav, A. Mohanty, J. Sussman, J. Lee, and M. A. Baldo, Nano Lett. 11, 1495 (2011).

[13] B. Ehrler, B. J. Walker, M. L. Boehm, M. W. B. Wilson, Y. Vaynzof, R. H. Friend, and N. C. Greenham, Nat. Commun. 3, 1019 (2012).

[14] D. N. Congreve *et al.*, Science 340, 334 (2013).

[15] J. R. Tritsch, W.-L. Chan, X. Wu, N. R. Monahan, and X. Y. Zhu, Nat. Commun. 4, 2679 (2013).

[16] J. J. Burdett, A. M. Mueller, D. Gosztola, and C. J. Bardeen, J. Chem. Phys. 133, 144506 (2010).

[17] P. M. Zimmerman, F. Bell, D. Casanova, and M. Head-Gordon, J. Am. Chem. Soc. 133, 19944 (2011).

[18] W.-L. Chan, M. Ligges, and X. Y. Zhu, Nat. Chem. 4, 840 (2012).

[19] M. W. B. Wilson, A. Rao, K. Johnson, S. Gelinas, R. di Pietro, J. Clark, and R. H. Friend, J. Am. Chem. Soc. 135, 16680 (2013).

[20] W.-L. Chan, M. Ligges, A. Jailaubekov, L. Kaake, L. Miaja-Avila, and X. Y. Zhu, Science 334, 1541 (2011).

[21] P. M. Zimmerman, Z. Zhang, and C. B. Musgrave, Nat. Chem. 2, 648 (2010).

[22] P. M. Zimmerman, C. B. Musgrave, and M. Head-Gordon, Acc. Chem. Res. 46, 1339 (2013).

[23] J. C. Johnson, A. J. Nozik, and J. Michl, Acc. Chem. Res. 46, 1290 (2013).

[24] J. J. Burdett, D. Gosztola, and C. J. Bardeen, J. Chem. Phys. 135, 214508 (2011).

[25] Z. Birech, M. Schwoerer, T. Schmeiler, J. Pflaum, and H. Schwoerer, J. Chem. Phys. 140, 114501 (2014).

[26] N. E. Geacintov, M. Binder, C. E. Swenberg, and M. Pope, Phys. Rev. B 12, 4113 (1975).

[27] S. Arnold and W. B. Whitten, J. Chem. Phys. 75, 1166 (1981).

[28] R. P. Groff, P. Avakian, and R. E. Merrifield, Phys. Rev. B 1, 815 (1970).

[29] C. Ramanan, A. L. Smeigh, J. E. Anthony, T. J. Marks, and M. R. Wasielewski, J. Am. Chem. Soc. 134, 386 (2012).

[30] M. J. Y. Tayebjee, R. G. C. R. Clady, and T. W. Schmidt, Phys. Chem. Chem. Phys. 15, 14797 (2013).

[31] J. J. Burdett and C. J. Bardeen, J. Am. Chem. Soc. 134, 8597 (2012).

[32] V. Coropceanu, M. Malagoli, D. A. da Silva Filho, N. E. Gruhn, T. G. Bill, and J. L. Brédas, Phys. Rev. Lett. 89, 275503 (2002).

[33] N. Vukmirović, C. Bruder, and V. M. Stojanović, Phys. Rev. Lett. 109, 126407 (2012).

[34] A. Girlando, L. Grisanti, M. Masino, I. Bilotti, A. Brillante, R. G. Della Valle, and E. Venuti, Phys. Rev. B 82, 035208 (2010).

[35] E. R. Henry, Biophys. J. 72, 652 (1997).

[36] O. Alter, P. O. Brown, and D. Botstein, Proc. Natl. Acad. Sci. USA 97, 10101 (2000).

[37]  G. Coslovich *et al.*, Phys. Rev. Lett. 110, 107003 (2013).

[38]  R. A. Kaindl, M. Woerner, T. Elsaesser, D. C. Smith, J. F. Ryan, G. A. Farnan, M. P. McCurry, and D. G. Walmsley, Science 287, 470 (2000).

[39]  The density of singlet exciton at ~ 3ns is over two orders lower than the maximum value from the TRFL trace. With such low density, the contribution of singlet exciton to TA data is very low that is below the resolution of the measurement.

[40]  Y. Zhang and S. R. Forrest, Phys. Rev. Lett. 108, 267404 (2012).

[41]  S. H. Lim, T. G. Bjorklund, F. C. Spano, and C. J. Bardeen, Phys. Rev. Lett. 92, 107402 (2004).

[42]  M. Voigt, A. Langner, P. Schouwink, J. M. Lupton, R. F. Mahrt, and M. Sokolowski, J. Chem. Phys. 127, 114705 (2007).

[43]  A. Camposeo, M. Polo, S. Tavazzi, L. Silvestri, P. Spearman, R. Cingolani, and D. Pisignano, Phys. Rev. B 81, 033306 (2010).

[44]  R. H. Dicke, Phys. Rev. 93, 99 (1954).

[45]  M. W. B. Wilson, A. Rao, J. Clark, R. S. S. Kumar, D. Brida, G. Cerullo, and R. H. Friend, J. Am. Chem. Soc. 133, 11830 (2011).

[46]  B. J. Walker, A. J. Musser, D. Beljonne, and R. H. Friend, Nat. Chem. 5, 1019 (2013).

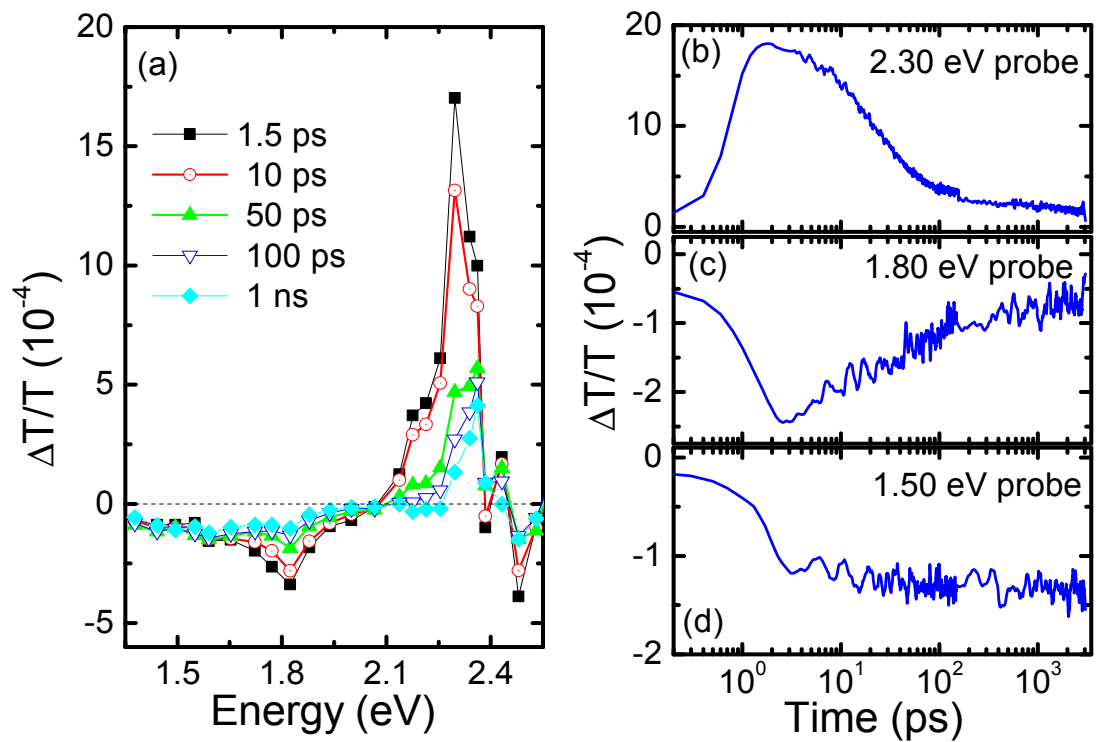

Figure 1, (a) The TA spectra of a tetracene film are shown at different pump-probe delays for photo-excitation at 3.1 eV. (b)-(d) The pump-probe traces probed at 2.30 eV, 1.80 eV and 1.55 eV with the pump at 3.1 eV, respectively.

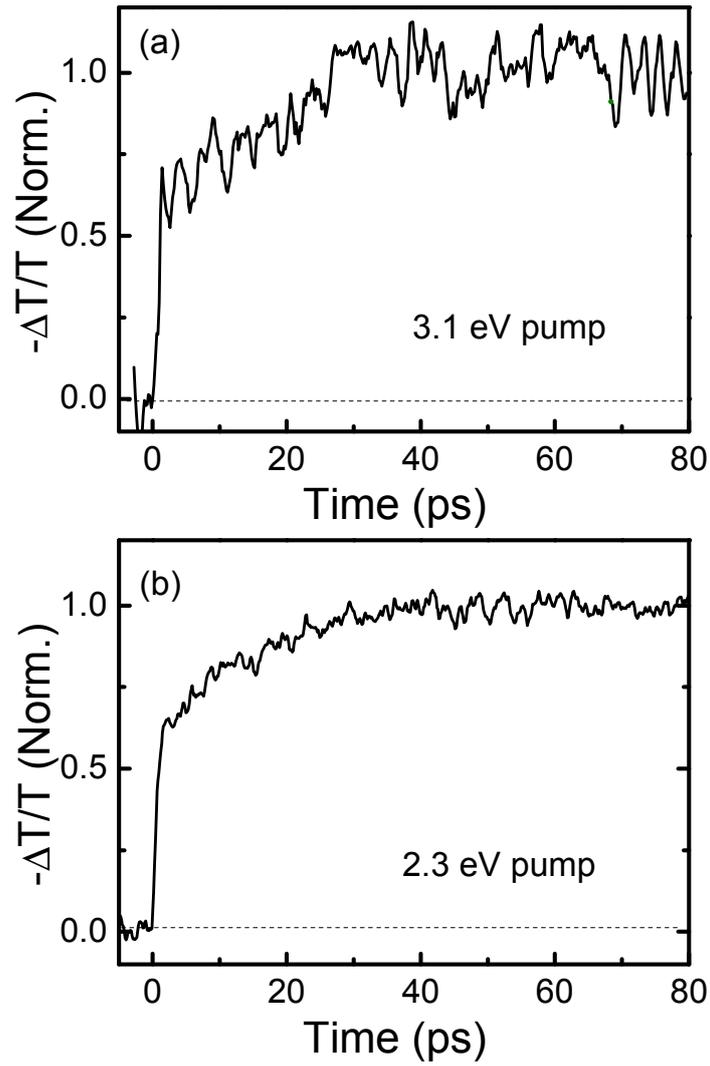

Figure 2, Normalized pump-probe spectra probed at 1.55 eV with the pump set at 3.1 eV (a) and 2.3 eV (b), respectively.

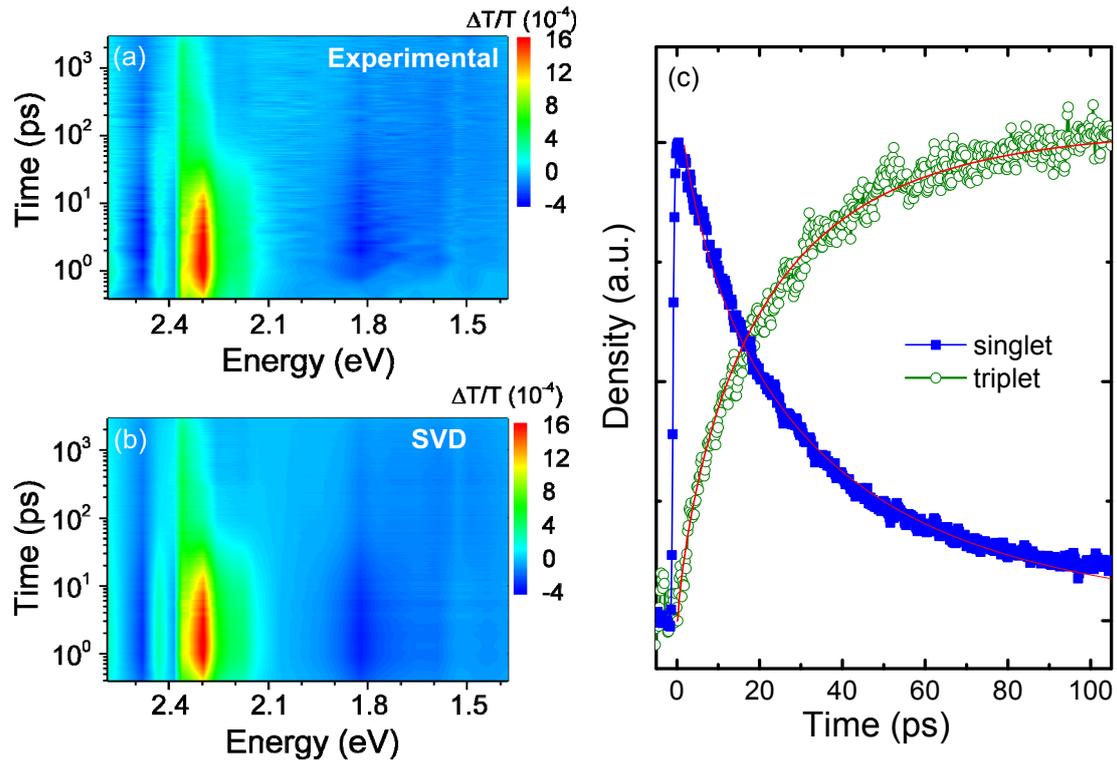

Fig.3, (a) Experimental data and (b) simulation results with SVD of time-energy matrices for photo-induced transmission change ($\Delta T / T$). (c) The dynamics of singlet and triplet populations extracted from the experimental data with the SVD method.

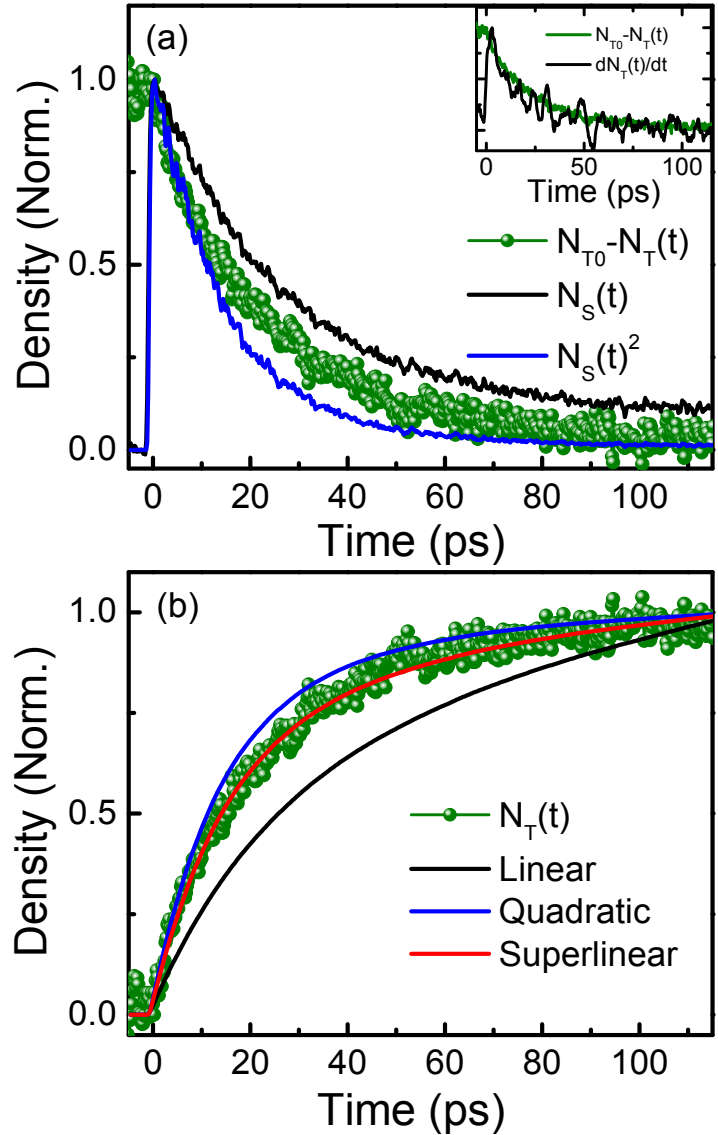

Fig. 4, (a) The temporal evolutions of $N_{T0} - N_T(t)$, $N_S(t)$, and $N_S(t)^2$ are compared with a normalized scale. Inset shows that $N_{T0} - N_T(t)$ is nearly proportional to the value of $dN_T(t)/dt$. (b) The build-up trace of triplet population is compared with the simulations based on linear, quadratic and superlinear density dependences in the integration form.

# Superlinear density dependence of singlet fission rate in tetracene films

(**Supplementary Information**)

## Sample preparation and characterizations

Tetracene powders were purchased from Sigma-Aldrich (sublimed grade, 99.99%). The films were prepared on glass substrates by thermal evaporation at a growth rate of ~ 0.1 nm/s with a base pressure of ~ $5 \times 10^{-5}$ Pa. Film thicknesses were calibrated with a surface profilometer (XP-2, Ambois). The thicknesses of the tetracene films for optical study are about 100 nm . The steady-state absorption and emission spectra as depicted in Fig. S1 show clear features of crystalline tetracene. The $S_1$ exciton states are characterized with two absorption peaks at 2.34 eV and 2.46 eV due to the Davydov splitting, and the absorption peaks at 2.63 eV and 2.81 eV are signatures of high singlet states ( $S_n$ ). The reference samples of single crystals were prepared with the method of physical vapor deposition in a quartz tube [S1]. The vapor of tetracene was carried by the argon flow (20 sccm) and then deposited onto substrates in the crystal growth zone. Single crystals with thickness of ~1 μm and size up to 5x5 mm$^2$ can be obtained. The crystalline quality was checked by the polarization microscopy and X-ray diffraction.

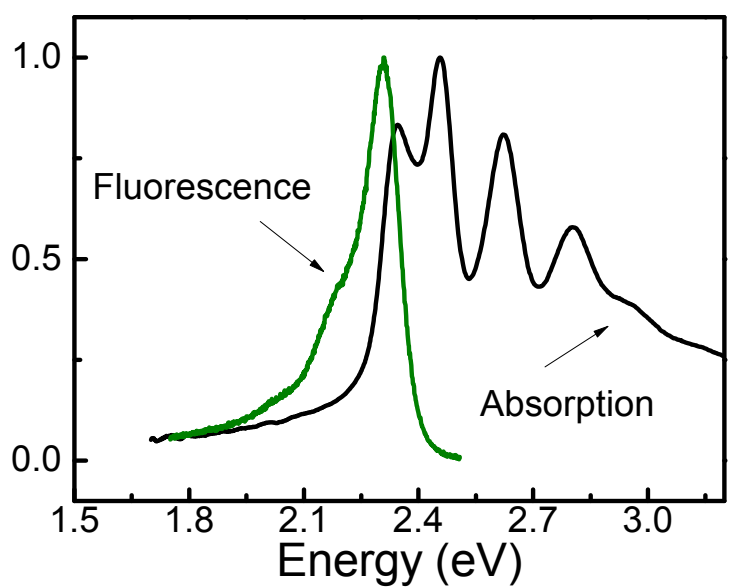

Fig. S1, Normalized absorption and fluorescence spectra of tetracene films.

We checked the fluorescence dynamics to confirm the presence of singlet fission in tetracene films. The time-resolved fluorescence (TRFL) spectra were recorded by the technique of time-correlated single-photon counting with a temporal resolution of ~ 50 ps as described previously [S2]. The quantum beats manifested as an oscillatory behavior in TRFL traces have been considered as a fingerprint of singlet exciton fission process [S3]. Figure S2 shows typical TRFL spectra recorded in tetracene solution, films, and single crystals, respectively. By subtracting off the multi-exponential decay components, we observed the damped oscillatory components in the TRFL traces in tetracene films and single crystals. The Fourier transformation of the oscillation components shows three peaks at 1.07, 1.84 and 2.95 GHz in the frequency domain. These values are in good agreement with the theoretical prediction of energy separations between manifold of the triplet-pair states [S3]. This result supports the presence of triplet pairs generated from singlet fission in crystalline tetracene.

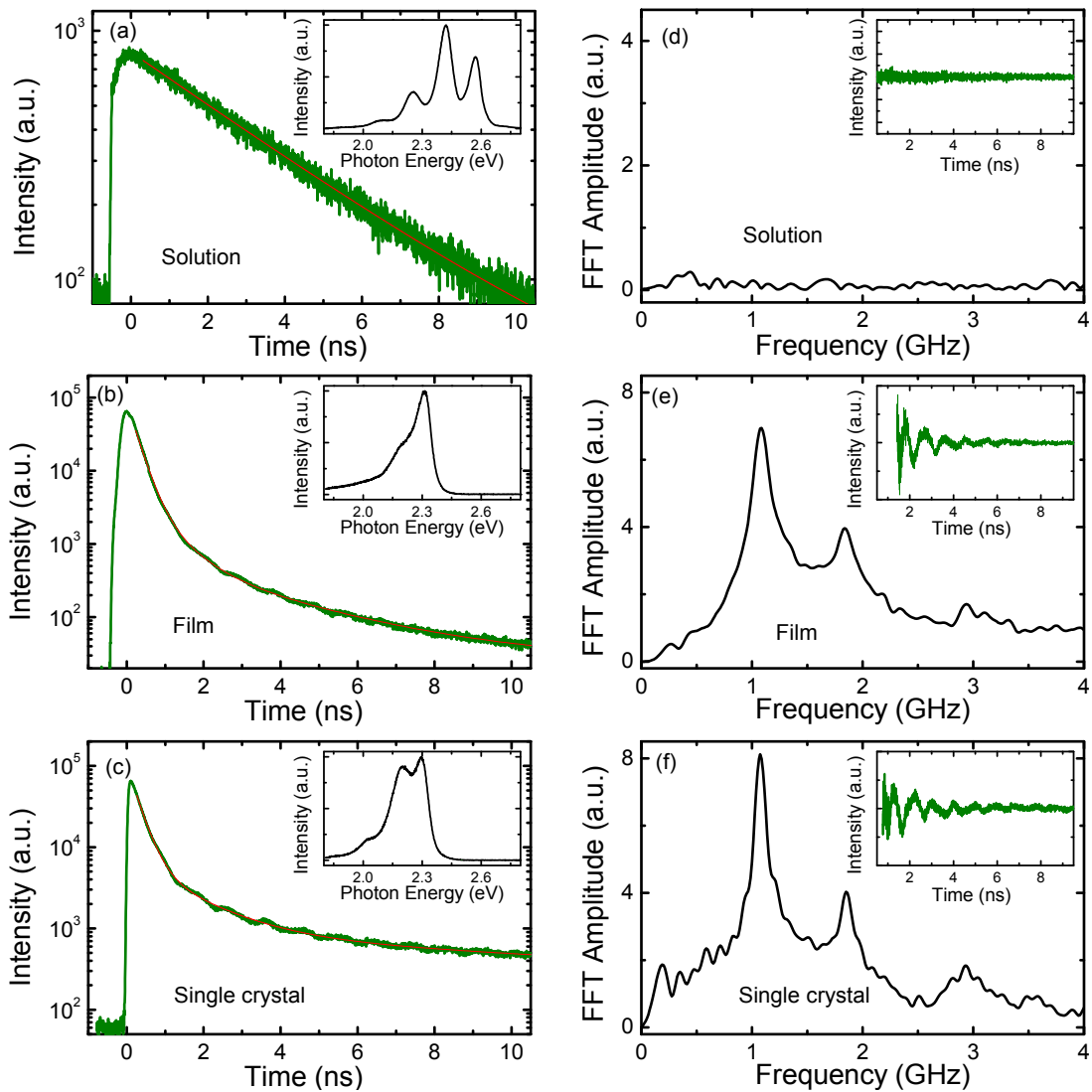

Fig. S2, (a)-(c), TRFL spectra recorded from the samples of tetracene solution, film and single crystal. Insets show normalized fluorescence spectra of the three samples, respectively. (d)-(e) The Fourier transform of the oscillation components in the three samples. Insets show the oscillations obtained by subtracting off the multi-exponential decay components in the TRFL spectra recorded in the three samples.

**Singular value decomposition (SVD) algorithm**

The TA data obtained in the pump-probe experiments can be regarded as a time-energy matrix $\Delta T / T(E, t)$. Previous analyses on the temporal dynamics for singlet and triplet excitons have been mainly done with single-wavelength data. However, due to the complicated structures of $S_n$ and $T_n$, the dynamic traces at single wavelengths may consist of multiple entangled components involving both singlet and triplet dynamics. We assume that the experimental data can be represented as a linear sum of independent components (i.e. $\Delta T / T(E, t) = \Sigma_{k=1}^{l} \phi_k(E) \varphi_k(t)$) and adopt the method of SVD to disentangle the spectrotemporal characteristics for singlet and triplet excitons. The method of SVD has succeeded in extracting the temporal and spatial information for complicated problems in several research areas [S4-7]. The details about the SVD technique are available in the literatures [S4-6]. In the following, we briefly review this method by considering a matrix of $A = \Delta T / T(E, t)$.

The SVD method decomposes the $M \times N$ matrix $A$ into a product of three matrices, i.e. $A = USV^T$, where $U$ is an $M \times M$ matrix with columns called left singular vectors $u_j$, $S$ is a diagonal matrix with elements $s_{jj}$ called singular values, and $V^T$ is an $N \times N$ matrix with elements of $v_j$ called right singular vectors. The singular vectors of $u_j$ and $v_j$ form orthonormal bases for the spectral ($\Delta T / T(E)$) and temporal ($\Delta T / T(t)$) traces. The singular vectors can be conventionally sorted by the amplitudes of singular values. The components with insignificant singular values are contributed basically by experimental noise, which can be naturally ignored.

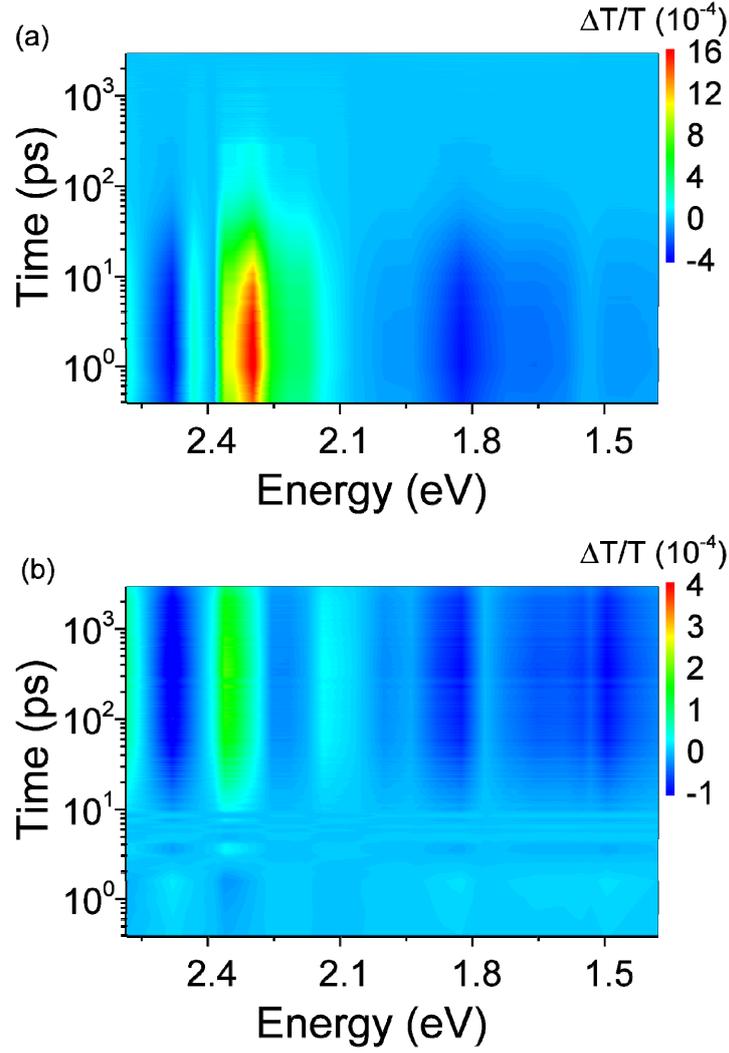

Fig.S3, The spectrotemporal features of singlet (a) and triplet (b) excitons extracted from the ultrafast TA data with the method of SVD.

The spectrotemporal characteristics for a physical process can be reconstructed by assuming a linear combination using the orthomormal bases, i.e. $\phi_k(\mathrm{E}) = \sum_j a_{kj} u_j$ and $\varphi_k(t) = \sum_j b_{kj} v_j$. In this study, the experimental data can be well reproduced with this SVD approach. The sum of $s_{11}$ and $s_{22}$ contributes over 90% of the trace of matrix S ($\sum_j s_{jj}$); namely the first two components contribute over 90% weight of the experimental data. To extract the temporal dynamics for the singlet and triplet excitons, we adopt the TA data at 0.5 ps and 3 ns as spectral features of singlet

( $\phi_1(E) = \Delta T / T(E, t = 0.5\,\text{ps})$ ) and triplet ( $\phi_2(E) = \Delta T / T(E, t = 3\,ns)$ ) populations. The decomposition procedure can then be simplified as a problem for a lower-rank matrix. The spectrotemporal features for singlet and triplet excitons are $\phi_1(E)\varphi_1(t)$ and $\phi_2(E)\varphi_2(t)$ as plotted in Fig.S3, respectively.

|        | $\tau_S$ (ps) | $\beta_S$ | $\tau_T$ (ps) | $\beta_T$ |
|--------|--------------|-----------|--------------|-----------|
| SVD    | 27.2         | 0.70      | 21.6         | 0.81      |
| 1.55 eV | 26.6        | 0.71      | 23.6         | 0.85      |
| 2.48 eV | 26.6        | 0.71      | 23.5         | 0.91      |

Table S1, The fitted parameters for the singlet and triplet dynamics derived by single-wavelength data (probed at 1.55 eV and 2.48 eV) and SVD, respectively.

For comparison, we follow the approach by Wilson *et al* [S8] and construct these temporal evolutions with single-wavelength data following Eq. 1. We choose the signals recorded at 1.55 eV and 2.48 eV with triplet features of long-lived photo-induced absorption. The probe photon energies are close to those used in previous studies [S8-11][??]. The temporal traces for triplet excitons are obtained by subtracting off the rapid rise component (i.e. the signal probed at ~ 2.2 eV) [S8] with normalization to the value at ~ 1 ps as shown in Fig. S4. For comparison, the phenomenological stretched-exponential (SE) equations in form of $N_S(t) = N_{S0} e^{-(t/\tau_S)^{\beta_S}}$ and $N_T(t) = N_{T0}(1 - e^{-(t/\tau_T)^{\beta_T}})$ are used to fit the decay and growth dynamics of singlet and triplet excitons, respectively. The fitted parameters, as listed in Table S1, show very good agreements with the dynamics derived from the

SVD method. With these results, we can safely exclude the rapid-rise component from the contribution of triplet excitons. The single-wavelength data consist of both singlet and triplet dynamics.

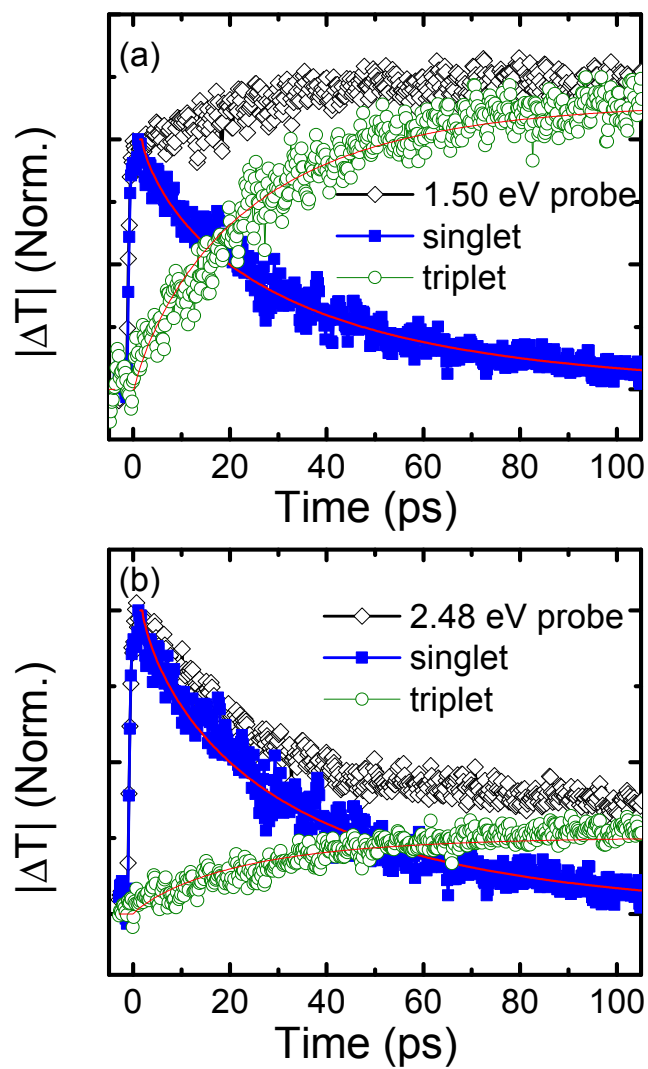

Fig.S4, The temporal dynamics for triplet excitons derived from the single-wavelength data with the probe-photon energy set at 1.55 eV (a) and 2.48 eV (b), respectively.

**Reference measurements with single crystals**

We also conducted ultrafast TA measurements on tetracene single crystals to check the divergent results observed in films and single crystals [S8-14]. The data are presented in Fig. S5. We have confirmed some differences discussed in literature: (1) The singlet excitons decay much slower in single crystals compared with that in films; (2) The ratio between amplitudes of triplet and singlet signals is much larger in single crystals. These results suggest that the trapping effect due to defects and grain boundaries expedites the relaxation of singlet excitons. The data can be well reproduced with the SVD method [Fig. S5(b)]. The reconstructed temporal traces for singlet and triplet excitons are plotted in Fig. S5(c). Their correlation shows a superlinear dependence of $k_{SF}$ on the singlet density. In comparison with the case of thin films, the fission rate $k_{SF}$ in single crystals is closer to the linear dependence, suggesting a more important role played by the spontaneous exciton fission.

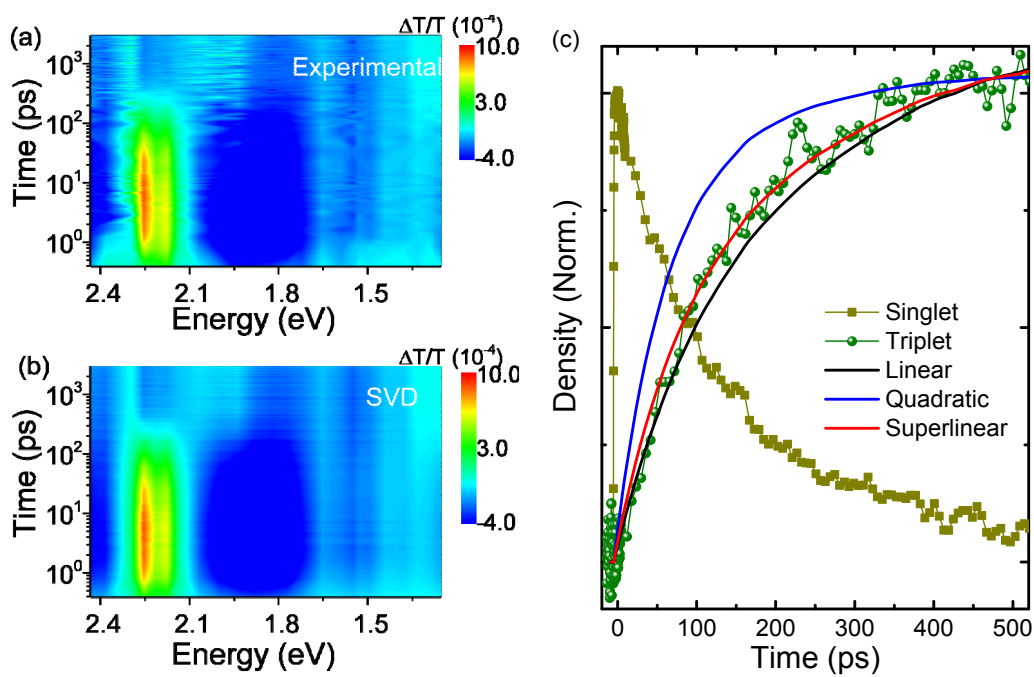

Fig.S5, (a) Experimental data and (b) simulation results with SVD of time-energy matrices for the transmission variation ($\Delta T / T$) recorded from tetracene single crystals. (c) The dynamics of singlet and triplet populations obtained with the SVD method. The build-up process is compared with the simulation data for linear, quadratic and superlinear dependent fission rates, respectively.

# References

[1]   V. Podzorov, Organic single crystals: Addressing the fundamentals of organic electronics, MRS Bull. 38, 15 (2013).

[2]   J. Xiao, Y. Wang, Z. Hua, X. Wang, C. Zhang, and M. Xiao, Carrier multiplication in semiconductor nanocrystals detected by energy transfer to organic dye molecules, Nat. Commun. 3, 1170 (2012).

[3]   J. J. Burdett and C. J. Bardeen, Quantum Beats in Crystalline Tetracene Delayed Fluorescence Due to Triplet Pair Coherences Produced by Direct Singlet Fission, J. Am. Chem. Soc. 134, 8597 (2012).

[4]   E. R. Henry, The use of matrix methods in the modeling of spectroscopic data sets, Biophys. J. 72, 652 (1997).

[5]   O. Alter, P. O. Brown, and D. Botstein, Singular value decomposition for genome-wide expression data processing and modeling, Proc. Natl. Acad. Sci. USA 97, 10101 (2000).

[6]   R. A. Kaindl, M. Woerner, T. Elsaesser, D. C. Smith, J. F. Ryan, G. A. Farnan, M. P. McCurry, and D. G. Walmsley, Ultrafast mid-infrared response of YRa2Cu3O7-delta, Science 287, 470 (2000).

[7]   G. Coslovich *et al.*, Competition Between the Pseudogap and Superconducting States of Bi2Sr2Ca0.92Y0.08Cu2O8+delta Single Crystals Revealed by Ultrafast Broadband Optical Reflectivity, Phys. Rev. Lett. 110, 107003 (2013).

[8]   M. W. B. Wilson, A. Rao, K. Johnson, S. Gelinas, R. di Pietro, J. Clark, and R. H. Friend, Temperature-Independent Singlet Exciton Fission in Tetracene, J. Am. Chem. Soc. 135, 16680 (2013).

[9]   J. J. Burdett, A. M. Mueller, D. Gosztola, and C. J. Bardeen, Excited state dynamics in solid and monomeric tetracene: The roles of superradiance and exciton fission, J. Chem. Phys. 133, 144506 (2010).

[10] E. M. Grumstrup, J. C. Johnson, and N. H. Damrauer, Enhanced Triplet Formation in Polycrystalline Tetracene Films by Femtosecond Optical-Pulse Shaping, Phys. Rev. Lett. 105, 257403 (2010).

[11] J. J. Burdett, D. Gosztola, and C. J. Bardeen, The dependence of singlet exciton relaxation on excitation density and temperature in polycrystalline tetracene thin films: Kinetic evidence for a dark intermediate state and implications for singlet fission, J. Chem. Phys. 135, 214508 (2011).

[12] V. K. Thorsmolle *et al.*, Morphology Effectively Controls Singlet-Triplet Exciton Relaxation and Charge Transport in Organic Semiconductors, Phys. Rev. Lett. 102, 017401 (2009).

[13] M. J. Y. Tayebjee, R. G. C. R. Clady, and T. W. Schmidt, The exciton dynamics in tetracene thin films, Phys. Chem. Chem. Phys. 15, 14797 (2013).

[14] Z. Birech, M. Schwoerer, T. Schmeiler, J. Pflaum, and H. Schwoerer, Ultrafast dynamics of excitons in tetracene single crystals, J. Chem. Phys. 140, 114501 (2014).